Discovery of periodic modulations in the optical spectra of galaxies, possibly due to ultra-rapid light bursts from their massive central black hole

Ermanno F. Borra,

Centre d'Optique, Photonique et Laser,

Département de Physique, Université Laval, Québec, Qc, Canada G1K 7P4

(email: borra@phy.ulaval.ca)

**SHORT TITLE:** Periodic modulations in the spectra of galaxies

**ABSTRACT:**

A Fourier transform analysis of 2.5 million spectra in the SDSS survey was carried out to detect periodic modulations contained in their intensity versus frequency spectrum. A statistically significant signal was found for 223 galaxies while the spectra of 0.9 million galaxies were observed. A plot of the periods as a function of redshift clearly shows that the effect is real without any doubt, because they are quantized at two base periods that increase with redshift in two very tight parallel linear relations. I suggest that it could be caused by light bursts separated by times of the order of $10^{-13}$ seconds because it was the original reason for searching for the spectral periodicity but other causes may be possible. As another possible cause, I investigate the hypothesis that the modulation is generated by the Fourier transform of spectral lines, concluding that it is not valid. Although the light bursts suggestion implies absurdly high temperatures, it is supported by the fact that the Crab pulsar also has extremely short unresolved pulses (<0.5 nanosecond) that also imply absurdly high temperatures. Furthermore, the radio spectrum of the Crab pulsar also has spectral bands similar to those that have been detected. Finally, decreasing the signal to noise threshold of detection gave results consistent with beamed signals having a small beam divergence, as expected from non-thermal sources that send a jet, like those seen in pulsars. Considering that galaxy centers contain massive black holes, exotic black hole physics may be responsible **for the spectral modulation**. However, at this stage, this is only a hypothesis to be confirmed with further work.

## 1. INTRODUCTION

The time domain is the least explored of all the physical astronomical domains (Fabian 2010). This is particularly the case for short timescales. The 2 nanosecond pulses, and the unresolved (< *0. 5* nanoseconds) nanopulses observed in the crab pulsar (Hankins et al. 2003, Hankins & Eilek 2007) are the shortest astrophysical time signals observed. Lorimer et al. (2007) reported a single 5 millisecond powerful burst of radio emission and searches for fast radio transients are presently carried out with VLBA data (Wayth et al. 2011). Astronomical objects that vary within times shorter than nanoseconds could be detected by searching for periodic modulations in their spectra (Borra 2010). Borra (2010) shows that objects that emit brief intensity pulses separated by times shorter than *$10^{-10}$* seconds induce spectral modulations detectable in astronomical spectra. The basic concept of the theoretical analysis in Borra (2010) can be intuitively understood by considering that the spectrum of a light source is given by the Fourier transform of the fluctuations of the electric field as a function of time (Klein & Furtak 1986). If we have two pulses of light separated by a time *t*, the spectrum will be modulated by a cosinus, because the Fourier transform of two separate peaks gives a cosinus. Note however that a spectral modulation does not have to be generated from only two pulses. It can be generated by a large number of pairs of pulses separated by the same constant time *t* but with each pair emitted with time separations much larger than *t* that could be periodic with a period much larger than *t*, or even emitted at random times much larger than *t* (Borra (2010). This is an important remark for a signal generated from only two pulses would imply an extremely large energy emitted in a very short time. The experiments of

Chin et al. (1992) support the theoretical analysis in Borra (2010). The present article discusses some of the results of a Fourier transform analysis of 2.5 million astronomical spectra in the Sloan Digital Sky Survey carried out to detect spectral modulations (Trottier 2012). Although the original motivation for the survey was to look for spectral modulations caused by ultra-rapid light bursts, spectral modulations could also be caused by other effects. For example, I consider in the present paper, two possibilities: instrumental effects and the effect of the Fourier transform of spectral lines, concluding that they do not apply. I do not know of any other possible cause.

## 2. DATA ANALYSIS

To search for the type of signal caused by intensity pulses having short time separations (Borra 2010), a Fourier transform analysis of 2.5 million spectra in the Sloan Digital Sky Survey (SDSS) was carried out to detect periodic modulations contained in their frequency spectra (Trottier 2012). The Fourier transform is carried out in the frequency domain which gives a Fourier spectrum in the time domain. This can be confusing since the language commonly used in textbooks is for Fourier transforms applied in the time domain, giving a Fourier spectrum in the frequency domain. Therefore **it is to be stressed that**, in this article, we will use the word "frequency" to refer to the units in the frequency spectrum before the Fourier transform. The units become time units after the Fourier transform of a frequency spectrum; however, in the discussion and the figures, we shall use the sampling number $N$, instead of time, so that the sampling can clearly be seen. Using time units would have been confusing in the discussion. The sampling number $N$ can be converted in time units (seconds) by multiplying it by  *2.1 10$^{-15}$*. The period in Hertz units in the frequency spectrum is therefore given by *1/(2.1 10$^{-15}$N )*.

The SDSS spectra are first converted to frequency units because the expected spectral modulation is periodic in frequency units (Borra 2010). This is done with the usual relations $v = c/\lambda$ and $\Delta v = c\Delta\lambda /\lambda^2$ . Furthermore, a Fourier transform carried out in the frequency domain requires that the spectrum be equally **sampled** in frequency. However, the SDSS spectra are equally sampled in wavelength so that they are not equally sampled in the frequency domain. Consequently a linear interpolation using the observed values on both sides of the required values, after conversion from wavelengths to frequencies, was used to generate equally sampled values in frequency.  The spectra are then analyzed using Fast Fourier Transform (FFT). Because the number of spectra is very large, simple signal finding algorithms must be used. A frequency spectrum is firstly smoothed with the matlab function *"smooth"* that uses a moving average filter. The smoothing length used is equal to 2.5% of the total length of the spectrum. This 2.5% length was chosen, after numerical experimenting, to ensure that it removes the short periods in the frequency spectra (corresponding to high values of $N$ in the Fourier domain) but does not remove the long periods (corresponding to low values of $N$). The smoothed spectrum is then subtracted from the unsmoothed one and the FFT performed on the difference between the unsmoothed and the smoothed spectra. This technique was used after experimenting with Fourier transforms of spectra without subtractions of the smoothed continuums and with subtractions of the smoothed continuum, including actual spectra and numerical simulations. The reason why the smoothed spectrum is subtracted is that, otherwise, there is a very strong and bumpy contribution at low values of $N$

(*N<40*) that would make it extremely difficult to detect a signal with software that analyzes an extremely large number of spectra (millions) and consequently must use simple numerical algorithms. In principle one could simply cut-off the region at low $N$; however, after trials and errors, it was found that subtracting the smoothed spectrum (and therefore low values of $N$ in the Fourier domain) gave better results. For example, smoothing in the frequency domain does not abruptly remove values of $N$ below a cut-off value but removes them gradually as a function of $N$ below a desired $N$ removal value. This allows one to see the Fourier spectrum even at low values of $N$ and to use simple numerical algorithms. This smooth transition with $N$, that occurs at $N < 40$, can be seen in figures 1, 2, 5 and 7 that show Fourier transforms of spectra. A frequency spectrum with subtraction of the smoothed spectrum is also easier to inspect visually (see figures 6 and 8). Comparisons of Fourier transforms of spectra with and without subtraction of the smoothed frequency spectrum, including numerical simulations, show that the subtraction does not remove signals for periods above the desired removal value and does not introduce artificial signals. An SDSS spectrum typically contains 3900 digital samplings in the frequency domain and therefore yields, after the FFT, 1950 samplings in the time domain. The FFT is carried out with the FFT function *"fft"* in Matlab software. The algorithms used by the function *"fft"* are explained in details in the help section of Matlab. In this paper we only consider the Fourier modulus which best characterizes the power of a signal and is always positive. Figure 1 shows the Fourier transform of the spectrum of a bright galaxy at RA = 50.53610 and Dec = -0.83626 (J2000) that has a strong signal. The signal at *N= 54* (corresponding to a time *t = 1.13 $10^{-13}$* seconds) is very sharp (only one sample), as expected after the Fourier transform of a periodic modulation that covers the entire spectral range observed. The Fourier modulus is plotted as a function of the sampling number *N*, instead of time, so that the sampling can clearly be seen. Beyond *N =500*, the *N* dependence of the Fourier modulus is essentially flat, within the noise, for it is dominated by white noise and shows the type and amplitude of fluctuations one sees for *400< N< 500* in Figure 1. This is the case for the majority of the SDSS objects. It is always the case for galaxies. In time units *N =1* corresponds to *2.1 $10^{-15}$* seconds and time increases linearly with *N*. Note that the total energy contained in the signal is actually not as strong as it appears, because most of the energy is contained in the subtracted smoothed spectrum. Also, the 3 arcseconds diameter of an SDSS fiber only samples a small fraction (the core) of this extended (about 1 arcminute) low-redshift (z = 0.0365) galaxy. To better display the shape of the signal, Figure 2 plots the first 100 samples of the Fourier transform in Figure 1. It clearly shows that the signal at *N = 54* is one sample wide. As discussed later, this is a very important characteristic of all of the signals detected.

To detect the type of signal seen in Figure 1 and 2, the software flags objects that have a peak, in the FFT spectrum, with a signal to noise ratio greater than a preset value. The signal is the signal of a single Fourier sample, the noise is evaluated with equation 2 and the signal to noise ratio is the ratio between these 2 numbers. Only a single sample is used by the software because its purpose is only to extract the Fourier transform spectra that have a statistically significant signal. Each extracted Fourier spectrum can then be visually inspected so that the presence of other significant samples can also be considered. The software is written in homemade Matlab code written to analyze SDSS spectra but it also uses Matlab functions: For example the functions *"smooth"* to carry out the smoothing and *"fft"* to carry out the fast Fourier transform. The software was extensively tested on SDSS spectra as well as numerical simulations. Because the modulus of the FFT is used, a statistical analysis must use Rayleigh statistics, which has the cumulative distribution function

$$F(x) = 1 - e^{-x^2/2\sigma^2} , \qquad (1)$$

where $\sigma$ is the standard deviation. To evaluate the noise, a Fourier spectrum is divided in 8 separate contiguous boxes of 250 $I_i$ samples and the standard deviation $\sigma$ is computed from the relation that must be used for Rayleigh statistics

$$\sigma^2 = \sum_{i=1}^{250} I_i^2 \ /500 \qquad (2)$$

for every box and this value of $\sigma$ is then used for all locations within that box. The box at the highest values of *N* has less than 250 samples (usually about 200) since 8 boxes of 250 samples cannot fit within the typical 1950 samples in the Fourier spectrum. No attempts are made to subtract the underlying "continuum" contribution, coming from the Fourier transform of the frequency spectrum, from the values of *Ii* used to compute the standard deviation in Equation 2 and to compute the signal. This is because, firstly, the underlying Fourier "continuum" is small for *N > 250*, even for bright objects, and contributes little to $I_i$ and therefore the signal to noise ratio. For *N > 500* the background signal is dominated by white noise for the majority of spectra and for all galaxies. For example the Fourier spectrum in Figure 1 continues, for *N > 500*, with a noisy appearance similar to what is seen for *400 < N < 500*. Secondly, for *N < 250* the continuum, which is negligible for faint objects (mostly galaxies) but significant for bright ones (mostly stars), must be evaluated from a bumpy Fourier modulus (like in Figures 1 and 2 but without the signal at *N = 54*) and depends on the spectral type of the object. It would therefore be extremely difficult to evaluate its contribution with software that runs without human intervention, as needed because of the huge quantity of spectra analyzed (2.5 millions). In the evaluation of $\sigma$ for *N < 250* and for bright objects, the underlying bumpy continuum signal therefore contributes to the noise evaluated from Equation 2, thereby decreasing the signal to noise ratio. On the other hand, the failure of removing the continuum has the opposite effect of adding a contribution, thereby increasing the signal. In practice, to avoid too large a number of detections that have to be visually inspected, only peaks giving a signal/to noise ratio > *6.5* had to be used for *N < 250* because there would otherwise have been too many detections of trivial signals to be visually inspected, coming from spectral bumps in very bright objects, mostly bright early type stars (see figure 5 for an example). Although the detection method has flaws for bright objects (mostly bright stars) and *N < 250* they are not important at this stage of the search, where the main purpose is to find whether peculiar objects exist and we are less concerned with a quantitative estimate of their occurrence rates. The software therefore simply flags interesting objects which can then be individually inspected. Note that these flaws are mostly important for stars, which tend to have bright spectra in the SDSS database, but not for galaxies because most of them are faint and, furthermore, the 3 arcseconds diameter of an SDSS fiber only samples a fraction of their apparent diameters: Consequently there is a small continuum contribution in the FFT spectra of the brightest galaxies even for *N< 250* (see Figure 1).

## 3. RESULTS

The analysis for *N > 250* gave several detections consistent with Rayleigh statistics. For *N < 250* a remarkable pattern, shown in Figure 3, is apparent for the signals detected in galaxies. The spectra of 0.9 million galaxies were analyzed (many galaxies were observed on different Julian Dates) but signals were detected in only 223 of them for a signal to noise ratio > 6.5. Figure 3 shows that the positions of the detected signals increase linearly with redshift along two linear staircases that have their bases (for a redshift $z = 0.0$) at $N = 52$ (*time* = $1.09 \cdot 10^{-13}$ seconds) and $N = 49$ (*time* = $1.03 \cdot 10^{-13}$ seconds). The staircases are consistent with a tight linear relationship sampled at the integer values of *N* given by the FFT. The staircase is caused by the effect of the discrete sampling in *N*. A linear relationship as a function of redshift signifies that the period of the spectral modulation is universal in the reference frame of the galaxies. On the one hand, this clearly shows that the signal is real and not an instrumental or data reduction artifact; on the other it places severe constraints on what causes it. The upper staircase relationship (with base at $N = 52$) shows that the signals are very sharp, like the signal seen in Figure 2. If the signals where broad (having width $\Delta N > 1$), the steps would not be so well defined. The staircase is therefore consistent with a spectral modulation, present in the entire observed spectrum of the galaxies that has the same value of *N* in their reference frame and increases linearly with redshift as observed on Earth. This effect can also be seen in the lower staircase, although it is not as well defined because of the smaller number of galaxies.

As discussed in section 2, the evaluation of the signal and of the noise for $N<250$ includes the contributions of the continuum. It was then stated that this does not have a significant effect. The statistical analysis that follows quantitatively elaborates on this. Firstly: Assuming that the periodic signal in Figures 1 and 2 is superposed to a background value equal to the continuum of the underlying Fourier spectrum, we find that the net signal has an amplitude of $I = 130$. Because the Fourier transform of white noise (photon noise in our case) also gives white noise, we can evaluate the standard deviation at large values of $N$, where the contribution from the Fourier continuum is totally negligible, and then apply it for an evaluation of the signal to noise ratio at $N = 54$. The white noise standard deviation is *8.0* for that spectrum, giving a signal to noise ratio of *16.25* for the signal and, using Equation 1, a probability of detection of a signal generated by random noise is $6.1 \ 10^{-34}$. Considering that $2.5 \ 10^6$ SDSS spectra were analyzed this gives a probability = $1.5 \ 10^{-27}$ that a signal generated by random noise be in a spectrum of an object at a given $N$ location. There are a few other objects that give comparable results. Secondly: Three of the galaxies in Figure 3 were detected with a signal to noise ratio $> 7.30$. A visual examination of their Fourier spectrum shows that the contribution $I$ of the underlying Fourier continuum of these faint galaxies is negligible so that the computed signal to noise ratio is reasonably accurate. Equation 1 gives a probability = $2.3 \ 10^{-12}$ that the signal is due to random noise. Considering that $2.5 \ 10^6$ SDSS spectra were analyzed this gives a probability = $5.8 \ 10^{-6}$ that a signal generated by random noise be in a spectrum of an object at a given $N$. The probability that 3 signals be detected from random noise is given by the product of the 3 probabilities and is therefore totally negligible.

To see the effect of lowering the signal to noise ratio selection minimum, a subsample of 128,000 spectra (5% of the total number) was analyzed again but with the threshold of detection of a signal lowered to a signal to noise ratio > *6.0*. The subsample came from all the spectra in the same region of the sky so that no selection effect is present. Signals were detected in 83 galaxies, 19 of which gave a signal to noise ratio > *6.5*. The positions of the detected signals as function of redshift are plotted in Figure 4. They give the same staircase relationships seen in Figure 3. Because the spectral subsample contains 5% of the total number of spectra analyzed we can estimate that lowering the signal to noise threshold to 6.0 for the entire survey would have yielded 1660 detections instead of 223, thus increasing the number of detections by a factor of 7.4. The fact that decreasing the signal to noise ratio of detection threshold of detection to 6.0 from 6.5 would only increase the number of detections by a factor of 7.4 while Rayleigh statistics (Equation 1) predict a factor of 23 is consistent with a very small fraction of galaxies that have strong signals, while the remainder have a substantially smaller signal or perhaps no signal all. Note that the majority of the detected galaxies, unlike the galaxy in Figures 1 and 2, are faint so that the continuum contributions to the signal to noise evaluations are small.

In conclusion, there is no doubt that a statistically significant signal has been found in a small number of galaxies. Only galaxies had this signal, while none of the stars and none of the quasars had statistically significant signals, and the signal was found in a very small fraction of the galaxies observed.

## 4. CONSIDERATION OF INSTRUMENTAL EFFECTS AND SIGNALS FROM THE FOURIER TRANSFORMS OF SPECTRAL LINES.

One must consider the possibility that the signals could be caused by instrumental effects. For example, the interference between two beams generated by reflections at two glass interfaces could generate the type of spectral modulations detected. However, because only a small number of galaxies were detected, while no stars nor quasars were, and the redshift relations in Figures 3 and 4, we can exclude that the peaks are due to instrumental effects. The next two paragraphs discuss the possibility of instrumental effects.

It would definitely be an instrumental effect if the detected objects were always located in the same fibers; however this is not the case for the detected objects are located in random fibers. The detected signals are weak, in comparison to the total energy in the spectrum, and an instrumental effect should therefore be prevalently detected in bright objects while this is not the case. Although the SDSS contains a large number of spectra of stars considerably brighter than the detected galaxies, the signal of the spectral modulation was not detected in any of the stellar spectra. Furthermore the detected galaxies are not particularly bright among the galaxies in the SDSS survey. As a matter of fact many of the detected galaxies are very faint so that a very large number of galaxies are brighter than the average detected galaxy; consequently, if the signal is instrumental, they should have the signal, but they do not.

The tight linear redshift relations seen in figures 3 and 4 also make an instrumental effect highly unlikely since an instrumental effect giving a periodic modulation in frequency units (e.g. interference between 2 optical beams) should be expected to give a period which is constant in the reference frame of the instrument and would not depend on redshift. The fact that there actually are two relations, which imply two signals having different periods at a redshift $z = 0.0$, makes an instrumental effect even more unlikely for one would expect both signals to be present in all detected objects, while an inspection of the Fourier spectra of all of the detected objects shows that they only have a single signal. Considering that the effect of the redshift is to shift to lower frequencies the energy distribution in the reference frame of the detector, a hypothetical instrumental effect should depend on the frequency dependence of the spectrum. However, the redshifts of the galaxies are small (half of the objects in Figures 3 and 4 have redshifts below 0.1) so that the effect of the redshift on the energy distributions is very small and unlikely to give such tight relationships with redshift. Furthermore there is a great diversity of spectral types among the detected galaxies. The effect should be more detectable in the bright galaxies but it is not. The fact that figures 3 and 4 show that the redshift dependence consists of two very close parallel lines makes a redshift-dependent instrumental effect even less likely since it would imply two quantized different spectral types, while a visual inspection of the galaxies in figures 3 and 4 showed a variety of spectral types (e.g. spectra of elliptical galaxies, emission-line galaxies, etc…) similar to the spectral types of undetected galaxies.

A staircase relation like the upper one in Figure 3 could occur if a peak centered at the required location was present in the Fourier spectrum of many galaxies. The small number of detections would then be due to the effect of noise fluctuations. An eye inspection of the Fourier transform of randomly chosen spectra of galaxies that were not detected, and of SDSS galactic templates, does not support this hypothesis. The Fourier spectra typically have bumpy structures like those seen in Figures 1 and 2 (without the signal at $N = 54$), but do not show evidence of higher peaks at the locations predicted by

Figure 3. Figure 7 shows the Fourier spectrum of an undetected galaxy. This hypothesis is also contradicted by the effect of decreasing the threshold of detection from 6.5 to 6.0 which increases the number of detections by a factor significantly smaller than predicted by Rayleigh statistics (see section 3)

There is a potential effect that could come from the Fourier transforms of spectral lines. The FFTs of the spectra of A and B stars that have strong absorption lines show a strong peak at $N = 37$ and a second weaker one at $N=57$. Figure 5 shows the Fourier transform of the spectrum of an A0 star. This is a typical Fourier spectrum one obtains from the FFT of A and B stars. It shows that the peak is broad (5 samplings) and that there is also a second weaker peak at $N = 57$. O stars also have a strong broad peak at $N = 45$ and a weaker peak at $N=62$. Computer simulations of artificial spectra that contain absorption lines show that the peaks are due the FFT of the contribution of the absorption lines. The simulations also show that the locations of the peaks vary linearly with redshift, leading to the suspicion that the relations in Figure 3 and 4 are due to the Fourier transform of spectral lines. However, the analyses that follow show that it is not the case. In reading the discussions one has to consider that the spectra of the SDSS have low spectral resolution (1800 to 2000 which corresponds to about 2.5 Angstroms) and low signal to noise ratios (Figures 6 and 8 show this for 2 bright galaxies); consequently, to be detected in a SDSS spectrum, spectral lines would have to be strong ones. A visual comparison of spectra of detected galaxies to spectra of hundreds of galaxies that were not detected did not show any noticeable differences, within the noise in the spectra, in their absorption or emission lines. The detected galaxies had unremarkable spectra. This is also the case of the spectrum of the galaxy that gave the strong signal seen in Figures 1 and 2. A visual inspection of the spectra of the detected objects in the SDSS web site shows that they also are, within the noise, well-fitted by SDSS spectrum templates and their spectral lines are identified as normal lines. Some have emission lines but the majority does not. The strengths of the emission lines vary greatly among the few detected galaxies with emission lines. There also is a diversity of galaxy types among the detected galaxies in Figure 3 (e.g. elliptical, emission line, star forming, etc...). Furthermore, the 3 arcseconds diameter of an SDSS fiber samples a large fraction of the diameters of the faint distant galaxies but only the nucleus of the nearby ones (e.g. the galaxy used to generate Figures 1 and 2). Note also that, unlike early type stars, the FFTs of later stellar spectral types, which are more representative of the spectra of detected galaxies, do not show significant peaks. In conclusion, on the basis of the eye inspection of the spectra, the detected galaxies do not have a striking similarity that makes them different from undetected galaxies.

Figure 6 shows the frequency spectrum of the galaxy of Figure 1 and 2 after subtraction of the smoothed continuum. The spectrum was blueshifted to a redshift *z = 0.0* to facilitate the comparison to Figures 8 and 11 later in the present paper. The spectrum in Figure 6 was smoothed with a box having a length of 3 times the spectral resolution in the SDSS spectrum to reduce the noise. This had to be done otherwise many of the weak lines would have been difficult to see by eye inspection in the background noise. This smoothing did not weaken significantly the spectral lines that were visible by eye inspection in the unsmoothed spectrum; therefore it should not weaken any line. Note that this smoothing was only done for eye inspection purposes only in Figures 6 and 8 and that the FFTs were always carried out with the unsmoothed spectra. Figure 7 shows the FFT of a galaxy ($z = 0.0465$) that was not detected. There clearly is no signal (expected at $N = 55$) like the one seen in Figures 1 and 2. Figure 8 shows the smoothed frequency spectrum of this galaxy after subtraction of the smoothed continuum. The spectrum was blueshifted to a redshift $z = 0.0$ to facilitate the comparison to Figures 6 and 11. The galaxy in Figure 6 has a redshift of 0.0365 while the galaxy in Figure 8 has a redshift of 0.0465 so that the spectral ranges displayed in both figures are nearly the same. The noise is significant in both figures. Also, in both spectra the noise increases significantly with frequency for frequencies higher than $7.0 \times 10^{14}$ Hz because the flux is smaller and the spectrograph less efficient at those frequencies. These objects were chosen for display purposes because they are among the brightest ones and their figures are easier to compare because fainter objects would have noisier spectra. Comparing the spectra of figure 6 and 8 we can see that they are very similar. Both spectra are well fitted by templates in the SDSS web site that clearly identifies the lines as common spectral lines (e.g. H, O, S, Ne lines) at the locations predicted by the redshifts. The relative strengths of the lines are not the same in the 2 spectra, partly due to random noise which is more important at the 2 frequency extremes. However, it is important to note that computer simulations show that the frequency location of the line is the fundamental factor to consider, for it determines the location of a peak after the Fourier transform, while varying strength only affects the strength of the peak. Consequently, because Figures 1 and 2 are obtained from the FFT of Figure 6 (without smoothing and without blueshifting) and Figure 7 from the FFT of Figure 8 (without smoothing and without blueshifting) and Figures 1 and 2 carry a strong signal while Figure 7 does not, it goes against the hypothesis that the signals are caused by the FFT of the spectral lines.

The similarities among the spectra of galaxies with a detected signal to the spectra of those that did not have a signal leads to the conclusion that the signal is not caused by the Fourier transform of spectral lines. The computer simulations discussed in the next section quantitatively elaborate on this.

## 6. COMPUTER SIMULATIONS

The discussion in the previous section concludes that the signals are not due to the Fourier transform of the spectral lines but it is mostly qualitative. This section presents the results of computer simulations that strengthen this conclusion.

The sharp peak seen in Figure 2 is typical of the peaks detected in the objects that are plotted in Figure 3. It looks like a Kronecker delta function. However, because it is the result of the Fourier transform of a spectrum, it cannot be a delta function because a delta function would be the result of the Fourier transform of the entire spectrum and be centered at $N=0$. It could however be the tooth of a comb-like function that we can conveniently model with a Shah function because the Fourier transform of a Shah function $III(f-nF)$ (where $n$ is an integer number and $F$ is the period of variation in frequency units) in the frequency domain $f$ gives another Shah function $III(t-n/F)$ having period $P = 1/F$ in the time domain $t$. We can therefore model the absorption lines in the frequency spectrum $I(f)$ of a galaxy by the convolution of a Shah function with a Gaussian function that conveniently models a line profile. The Fourier transform of a convolution is the product of the Fourier transforms of the 2 functions. The Fourier transform of a Gaussian having dispersion $\sigma$ is another Gaussian having dispersion $\sim 1/\sigma$. Consequently the Fourier transform of the spectrum $I(f)$ having periodically recurrent spectral lines, modeled by a Shah function *III(f-nF)* with period *F* convolved with a Gaussian function having dispersion $\sigma$, is another Shah function *III(t-n/F)* having a period *P = 1/F* multiplied by a Gaussian having dispersion $\sim 1/\sigma$ centered at *N=0*.

Starting from the analytical discussion in the previous paragraph, I have carried out numerical simulations, using matlab software, that model an SDSS spectrum in the frequency domain with a flat continuum to which one subtracts the convolution of the Shah function with a Gaussian to simulate a spectrum that contains absorption lines. Because the time shown in Figure 2 is $T = 1.13 \: 10^{-13}$ seconds and the Fourier transform of the Shah function gives a frequency period $F = 1/T = 8.45 \: 10^{12}$ Hz, the frequency spectrum must contain spectral lines separated by $8.45 \: 10^{12}$ Hz. Figure 9 shows the result of the Fast Fourier Transform (FFT) of a spectrum that uses this model. The dispersion of the Gaussian function used to simulate the absorption lines was obtained from the average width of the absorption lines in the spectrum of the galaxies in Figures 6 and 8, which is typical of the width of the lines in a typical galaxy observed in the SDSS survey. The fact that we approximate the spectral line of a galaxy with a Gaussian only causes the minor error that the Shah function *III(t-n/F)* is multiplied by the Fourier transform of a Gaussian instead of the Fourier transform of the exact shape of the spectral line. Because both these functions have the same half-widths we can understand that the approximation has a minor effect. Figure 9 confirms the theoretical discussion based on the Shah function in the previous paragraph. However, it obviously is not in agreement with Figure 1 since it only shows one peak, while on the basis of Figure 9 we would have expected 3 visible peaks, even considering the presence of noise. Furthermore, the numerical frequency spectrum that generates Figure 9 contains 50 spectral lines equally spaced by

$8.45 \times 10^{12}$ *Hz* and having equal peak intensities = -0.6 in the smoothed spectrum, which is in obvious disagreement with the spectrum in Figure 6.

To understand the effect of changing the intensities of the absorption lines as well as their frequency locations more simulations were carried out. They used the same basic spectrum of the convolution of the Shah function with a Gaussian that generated Figure 9, but added matlab code that changed at random the intensities of the lines and the frequency locations of the individual lines. The intensity of the lines was changed by multiplying the peak intensity of every line with the matlab function "*rand*" that generates random numbers having values uniformly distributed between 0 and 1. The frequency locations of the absorption lines were also changed with *rand* multiplied by a constant smaller than one and by the equal frequency separation of $8.45 \times 10^{12}$ *Hz* in the Shah function. Such a random frequency location shift generated by *rand* was then added to the frequency position of the Shah function. A different frequency deviation from the periodic frequency location predicted by the Shah function was therefore randomly generated for every one of the periodic 50 frequency locations of the absorption lines. After several simulations, it was found that a signal similar to the signal seen in Figure 2 could be generated with appropriate values of the constant that multiplies *rand*. Figure 10 shows an example of such a simulation. A strong peak, similar to the one in Figure 2 can clearly be seen. At first sight this seems to confirm the hypothesis; however, in practice it does not because too many strong lines at the appropriate peculiar frequency locations are required. I elaborate on this in the next paragraphs.

Figure 11 shows the frequency spectrum that was used to generate Figure 10 with a Fourier transform. The intensities of the lines were chosen to give a signal equal to the signal seen in Figures 1 and 2; consequently, this spectrum can be readily compared to the spectrum in Figure 6 as well as the spectrum in Figure 8. The positions of the lines have the periodicity (in frequency units) required to give the peak seen in Figure 10, but have an additional random deviation from that value generated with the function *rand* (see previous paragraph). The average deviation from the *8.45 $10^{12}$ Hz* frequency period is +- *7.5%* , which corresponds to a +- *0.1 %* deviation from the periodic frequency position, so that the positions of the spectral lines are still reasonably periodic in frequency. A comparison of Figure 11 to Figures 6 and 8 shows that a large number of absorption lines strong enough to be seen by eye inspection should be present in Figure 6 but not in Figure 8. Of course one does not need to add 50 spectral lines to the spectrum of Figure 8 to generate, after the Fourier transform, a signal similar to the signal seen in Figure 2. Firstly some of the lines in Figure 11 are weak and would not be detectable by eye inspection in Figures 6 and 8. Secondly, because one would need to add lines to the spectrum of Figure 8, some of the lines in that spectrum would presumably already be near the needed location and one would therefore have to add a number smaller than 50. Note that Figure 8 could only have a very small number of lines at the required frequencies; otherwise they would have generated a visible peak in Figure 7. A comparison of Figures 6, 8 and 11 therefore clearly shows that significant numbers of lines detectable by eye inspection would have to be added to Figure 8 to generate a peak similar to the peak visible in Figure 2. Half of the lines in Figure 11 have an intensity equal or deeper than -0.4 and therefore should be easily detectable in Figures 6 and 8. In conclusion, because Figure 6 and 8 are so similar, we can exclude that the required number of additional lines is present in Figure 6.

Note that the total number of required strong lines seen in the theoretical Figure 11 is about the same total number of lines visible in the spectra in Figures 6 and 8. Finally, the required lines must be present in a continuous spectral region that is at least 70 % of the total spectrum, otherwise, after the Fourier transform, the peak would be broader than what is seen in Figure 10 with a width inversely proportional to the width of the region that contains the lines. For example, if the periodic lines are only present in the left half of the spectrum (between *320 THz* and *550 THz*) and not the right half, the peak in Figure 2 would have been 2 samples wide ($\Delta N = 2$) instead of one ($\Delta N = 1$). If the lines are only present in a region having a width equal to 1/4 of the spectrum, the peak would have been 4 samples wide. ($\Delta N = 4$).

The hypothetical lines that would generate the signal must be among the strongest ones in the spectra. This can be seen by comparing the numerical simulation figures to the observational figures. This can also be seen in the Fourier transform domain. Figure 10, generated by the FFT of Figure 11, shows that the average Fourier "continuum" generated by the FFT of the lines is strong because it has an amplitude of 50, half the amplitude of 100 of the Fourier "continuum" of the detected galaxy in Figure 1. However, the numerical Fourier spectrum in Figure 10 does not include the effect of random noise. The standard deviation due to random noise in Figures 1 and 2 is equal to 8; therefore, to estimate the contribution of noise one must add to the numerical Figure 10 a noise contribution with an upper limit of about 24, equal to 3 standard deviations. This gives an amplitude of 80, comparable to the Fourier continuum amplitude in Figure 2.

## 7. DISCUSSION AND CONCLUSION

A statistically significant signal has been found in a very small fraction of the galaxies observed (223 out of 0.9 millions) for a signal to noise ratio $> 6.5$. No signal was detected at these $N$ locations in over 0.5 million stars and quasars observed. All evidence is consistent with a signal coming from a periodic modulation of the frequency spectrum. The tight redshift dependences shown in Figures 3 and 4 and the fact that neither a single star nor quasar was detected exclude the possibility of instrumental effects. Computer simulations show that the signal shown in Figures 1 and 2 implies a periodic modulation, like in Equation 5 in Borra (2010), with amplitude +- 0.2 in its frequency spectrum in Figure 6 so that the signal is buried in noise in that figure and it is normal that it does not show up by eye inspection.

The hypothesis that the signal is caused by the Fourier transform of spectral lines, discussed in sections 5 and 6, is not in agreement with the observational evidence because the spectra of galaxies that have the signal and the spectra of those that do not have it are too similar. In reading the discussions in section 5 and 6, one has to consider that the SDSS spectra of the galaxies have low spectral resolution and low signal to noise ratios, consequently, to give a signal after the FFT of a SDSS spectrum, spectral lines would have to be strong ones. Besides the observational evidence, other arguments go against the hypothesis. In principle large differences in stellar type distributions may do the job. However, there is no justifiable reason that adding different spectral types would give the needed periodic frequency modulation. The FFTs of the spectra of late spectral type stars, which are more representative of the spectra of galaxies do not show any significant peaks at all. The only stars that have significant peaks are early type stars and their peaks are not located at $N = 52$ as required. Note also that the detected galaxies have a single peak (see Figures 1 and 2), while the FFTs of A and B stars have two broad peaks with the main peak at *N = 37* and a weaker peak at *N = 57* (see Figure 5) while O stars have the main peak at *N = 45* and a weaker peak at *N=62*. Adding more lines at the required frequencies would narrow the widths of the Fourier peaks but not change their locations. Another possibility comes from differences in chemical abundances. However, there is no obvious evidence that changing chemical compositions would give additional lines at the required spectral locations. Because the hypothetical lines are absorption lines in the spectra of galaxies it would imply large peculiar chemical compositions in a large fraction of their stars. Intuitively, the hypothesis of highly peculiar chemical abundance may be credible for a single star but, on the basis of what we know about chemical abundances in the universe, it seems highly unlikely that the large fraction of stars needed in a galaxy would have strong very peculiar chemical compositions. Note also that simply increasing the known chemical compositions of galaxies would not do the job. This is because the Fourier spectra of the undetected galaxies do not show any evidence of a weak peak (see Figure 7). Furthermore the number of detected galaxies is very small and decreasing the signal to noise ratio criterion increases the number of detections below the number expected from Rayleigh statistics (see section 3). Another major problem with the FFT of spectral line hypothesis comes from the fact that, in Figures 3 and 4, the positions of the signals increase linearly with redshift along two linear staircases that have different bases (for a redshift *z = 0.0*) at *N = 52* ($1.09 \ 10^{-13}$ seconds) and *N = 49* ($1.03 \ 10^{-13}$ seconds). Consequently one would need two sets of galaxies having spectral lines positioned at two slightly different frequency periods and therefore two different types of highly peculiar chemical abundances. This renders the hypothesis even less credible, especially considering that we are dealing with the spectra of galaxies and not with the spectra of single stars. It must be noted that even if turns out that the signal is due to the Fourier transform of the spectral lines there would be an interesting discovery because it would imply the existence of a small number of galaxies with highly peculiar chemical compositions. An obvious question that arises is: What is the fundamental reason why spectral lines in the simulations generate a signal similar to a cosinusoidal modulation? The answer is: The signal that is added comes from spectral lines that are positioned at periodically recurrent positions so that we can model it with a cosinusoidal curve.

At this stage, one can only speculate on the physics responsible for the spectral modulation. This is done in the following paragraphs.

As discussed in section 3, decreasing the signal to noise threshold to 6.0 from 6.5 for the entire survey would have increased the number of detected galaxies by a factor of 7.4 (from 223 to 1660) while Rayleigh statistics predict a factor of 23. The fact that the detections are below those predicted by Rayleigh statistics, which apply to our detections because we use the Fourier modulus, by a significant factor indicates that the signals are strong signals that come from a very small fraction of galaxies. This is consistent with signals coming from strong light beams present in many galaxies but detected in only a few of them because the beams have small divergences. This is what one intuitively expects from non-thermal sources that send a jet. This is also the kind of beam seen in pulsars.

The signal could be generated by sources sending pulses separated by times of the order of $10^{-13}$ seconds, which was the original principal reason for searching for spectral periodicities (Borra 2010) but, at this stage, this is only a hypothesis to be verified with further work. One may object that to generate pulses separated by $10^{-13}$ seconds would necessitate very small volumes having absurdly large implied brightness temperatures. As an answer to this objection one can note that the extreme unresolved (< *0.5* nanoseconds) pulses of the crab pulsar have a similar problem resulting in an implied brightness temperature of $2\ 10^{41}\ K$ (Hankins & Eilek 2007). Borra (2010) elaborates on this. Furthermore, the radio spectrum of the crab pulsar also shows spectral bands (Hankins & Eilek 2007) that have the type of spectral modulation that has been detected in galaxies, although the spacing between the radio bands increases linearly with frequency instead of being constant. This may possibly due to the effect of the pulsar environment since electromagnetic waves at radiofrequencies are affected by high-energy ionized environments they flow through.

While, on the one hand, the strong relationship seen in Figures 3 and 4, which implies that the period is constant in the reference frame of the galaxies, clearly indicates the presence of a puzzling effect; on the other, it places severe constraints on the Physics responsible. Because galaxies have a central massive black hole, unknown exotic physics may be responsible. Other explanations are also obviously possible.

Some galaxies were observed on different Julian dates. Some were detected (e.g. the galaxy in Figure 1), albeit with a lower signal to noise ratio, but some were not. This may be due to variability but also the effect of random noise.

The detected objects shall have to be observed again to confirm the effect in spectra with high signal to noise ratios. A list of the objects can be found in Trottier (2012) and will be published in a forthcoming paper. I will also send it upon request.

ACKNOWLEDGEMENTS

This research has been supported by the Natural Sciences and Engineering Research Council of Canada. The author wishes to thank Eric Trottier for the remarkable work he performed by analysing the SDSS data. Funding for SDSS-III has been provided by the Alfred P. Sloan Foundation, the Participating Institutions, the National Science Foundation, and the U.S. Department of Energy Office of Science. The SDSS-III web site is http://www.sdss3.org/.

REFERENCES

Borra, E.F. 2010, A&A letters 511, L6.

Chin, S.L., Francois, V., Watson, J.M., & Delisle, C. 1992, Applied Optics, 31, 3383.

Fabian, A.C.,2010, Serendipity in Astronomy In: Serendipity. The Darwin College Lectures", eds. M. de Rond & I. M. de Rond and I. Morley, Cambridge UP.

Hankins, T. H. , Kern, J. S. , Weatherall, J. C. & Eilek , J. A. 2003, Nature 422, 141

Hankins, T.H. & Eilek, J.A., 2007, ApJ 670, 693.

Klein, M. V., & Furtak, T. E. 1986 (New York: Optics, John Wiley & Sons)

Lorimer, D.R., Bailes, M., McLaughlin, M.A., Narkevic, D.J., & Crawford, F. 2007, Science 318 777.

Trottier, E., 2012, Recherche de signaux périodiques dans des spectres astronomiques, M.Sc. Thesis, Université Laval.

Wayth, R.G. et al. 2011, ApJ 735, 97.

Figure Captions

Figure 1

Fourier transform of the frequency spectrum (after subtraction of its smoothed frequency spectrum) of the core of a bright galaxy with a strong signal. The FFT modulus is plotted as a function of the FFT sampling number $N$. Only the first 500 samples (out of 1950) are plotted so that the signal at $N= 54$ ($1.13\ 10^{-13}$ seconds in time units) can clearly be seen. In time units $N =1$ corresponds to $2.1\ 10^{-15}$ seconds and time increases linearly with $N$. The subtraction of the smoothed spectrum removes the strong contribution present in the FFT of the unsmoothed spectrum for $N < 20$ (see section 2).

Figure 2

First 100 samples of the Fourier transform spectrum in Figure 1. The signal at $N= 54$ ($1.13\ 10^{-13}$ seconds) is very sharp (only one sample), as expected after the Fourier transform of a periodic spectral modulation that extends at least over the entire SDSS spectrum

Figure 3

Signals detected in galaxies as a function of redshift. The positions of the signals increase linearly with redshift along two linear staircases that have their bases (for a redshift $z = 0.0$) at $N = 52$ ($1.09\ 10^{-13}$ seconds) and $N = 49$ ($1.03\ 10^{-13}$ seconds).

Figure 4

Positions of the signals detected in galaxies as a function of redshift for a subsample of 128,000 spectra that was analyzed again but with the threshold of detection of a signal lowered to a signal to noise ratio $> 6.0$.

Figure 5

Fourier transform of the frequency spectrum (after subtraction of its smoothed frequency spectrum) of a bright A0 star. Only the first 100 values of $N$ are plotted.

Figure 6

Spectrum of the galaxy of Figure 1 and 2, in frequency units, after subtraction of the smoothed continuum. The spectrum was blueshifted to a redshift $z = 0.0$ to facilitate the comparison to Figures 8 and 11. The amplitude of the detected periodic signal is+- 0.2 in the frequency spectrum so that the signal is buried in noise and it is normal that it does not show up by eye inspection in this figure.

Figure 7

Fourier transform of the frequency spectrum (after subtraction of its smoothed frequency spectrum) of a typical galaxy that did not give a signal.

Figure 8

Spectrum of the galaxy in Figure 7, in frequency units, after subtraction of the smoothed continuum. The spectrum was blueshifted to a redshift z = 0.0 to facilitate the comparison to Figures 6 and 11.

Figure 9

Fast Fourier Transform of a Shah function convolved with a Gaussian.

Figure 10

Fast Fourier Transform of a simulation that used the same basic spectrum of the convolution of the Shah function with a Gaussian that generated Figure 9, but with computer code that changed at random the intensities of the lines (from 0.0 to -0.7) and the frequency locations of the individual lines (+- *0.1 %* average deviation from the Shah function frequency positions).

Figure 11

Frequency spectrum used to generate Figure 10 with a Fourier transform.

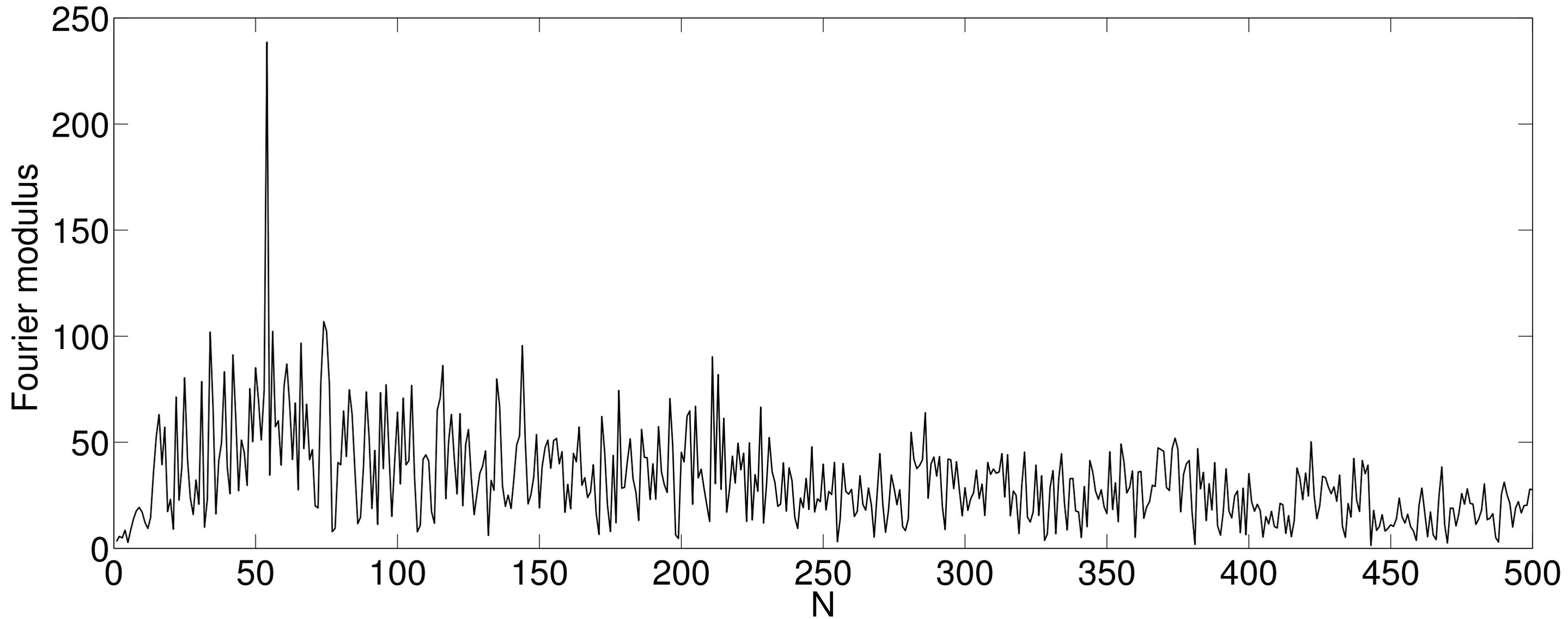

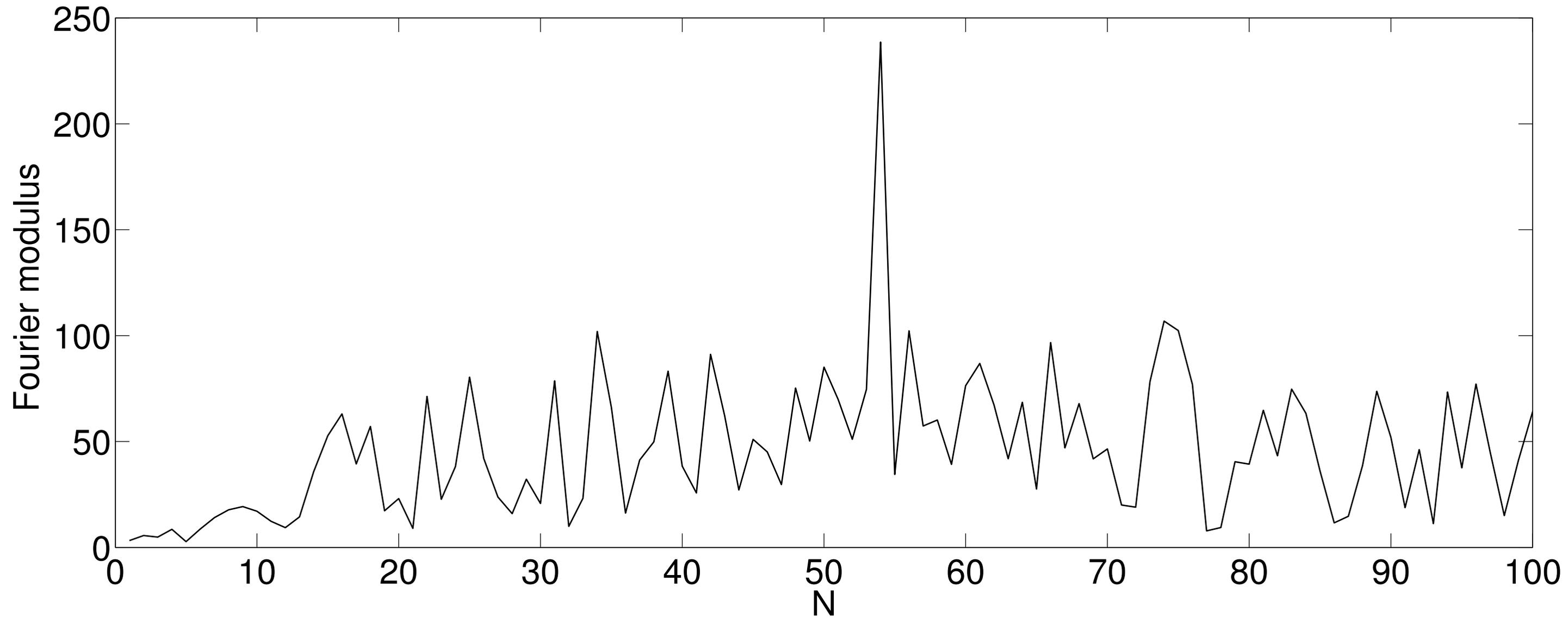

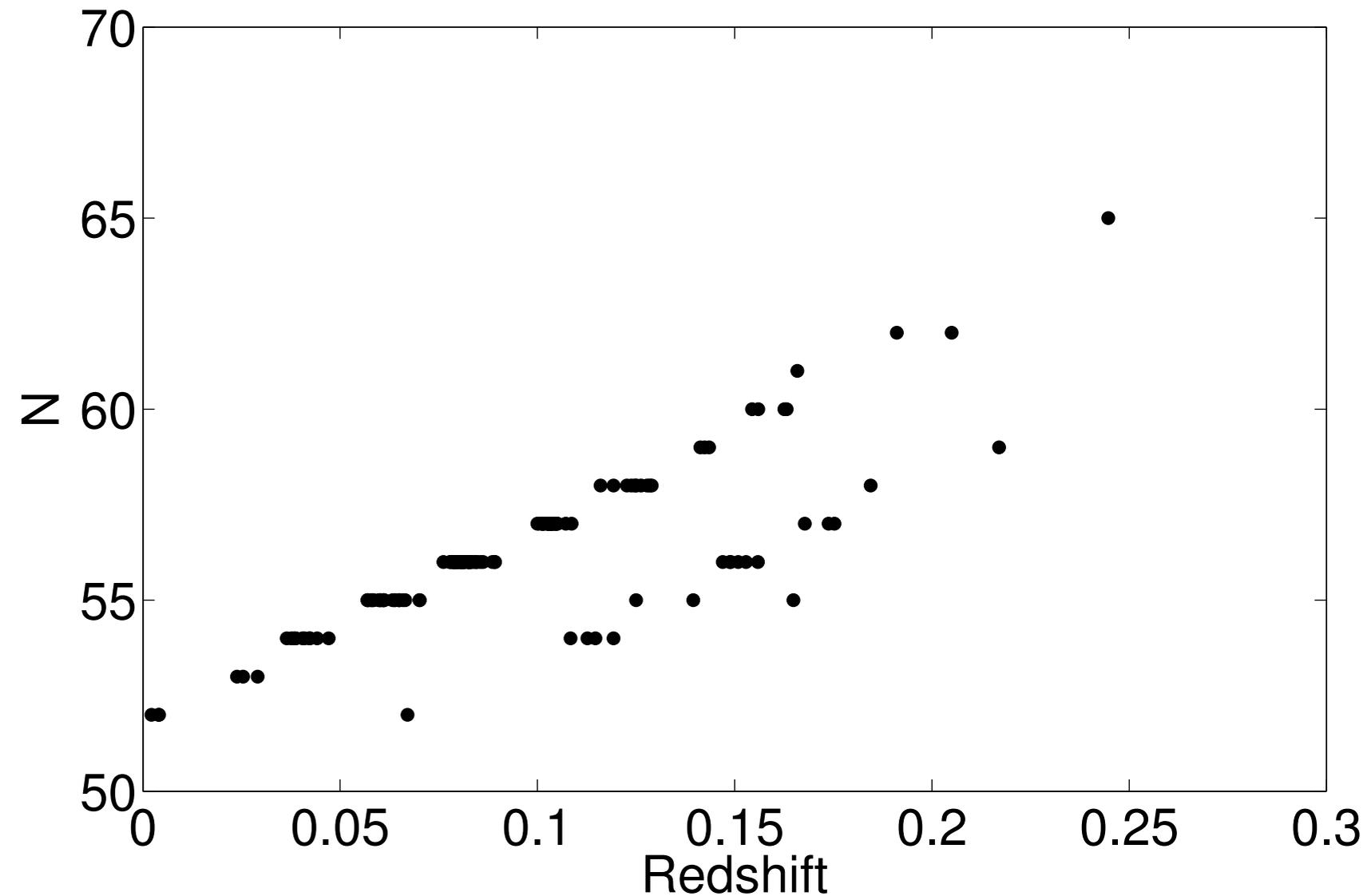

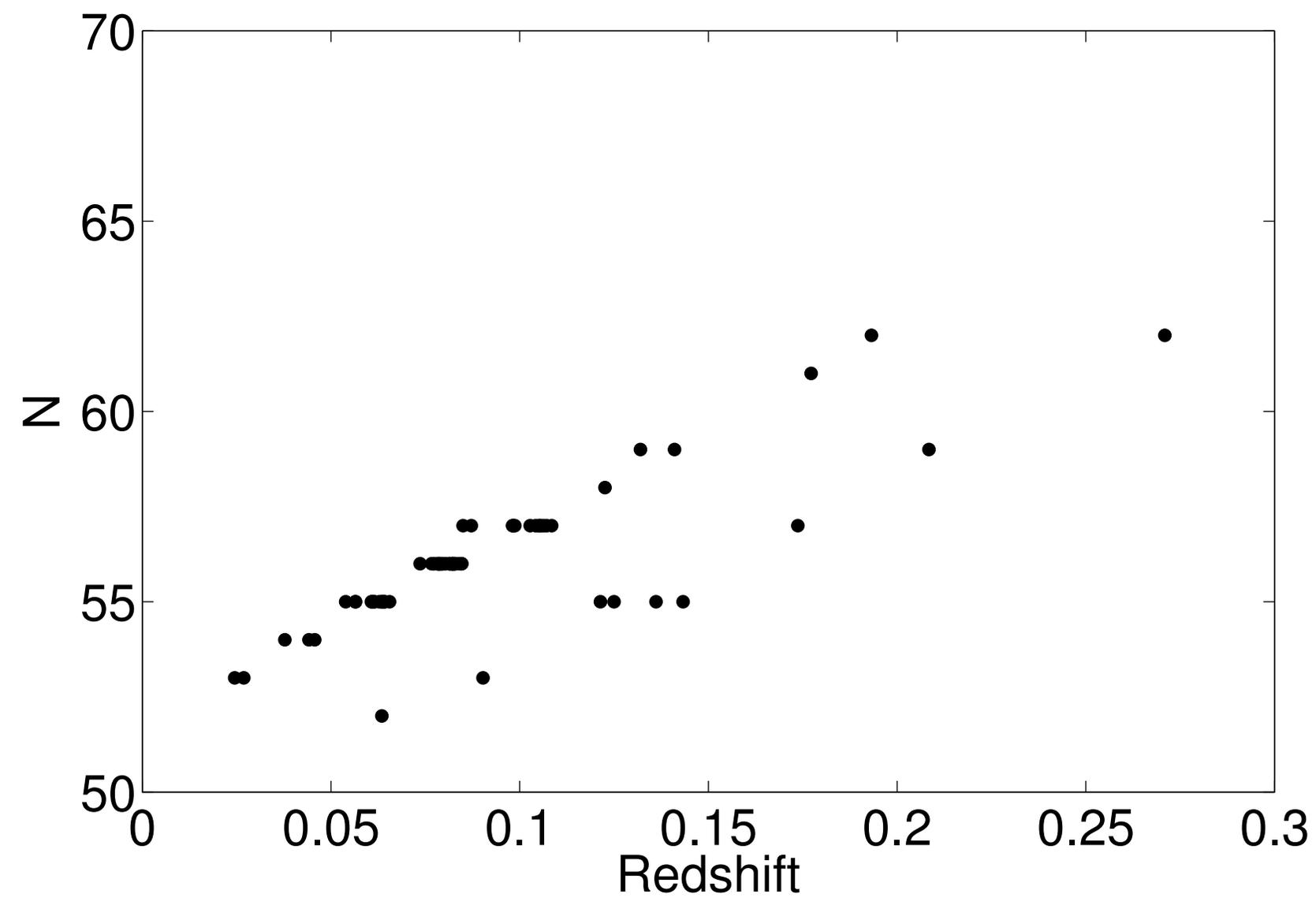

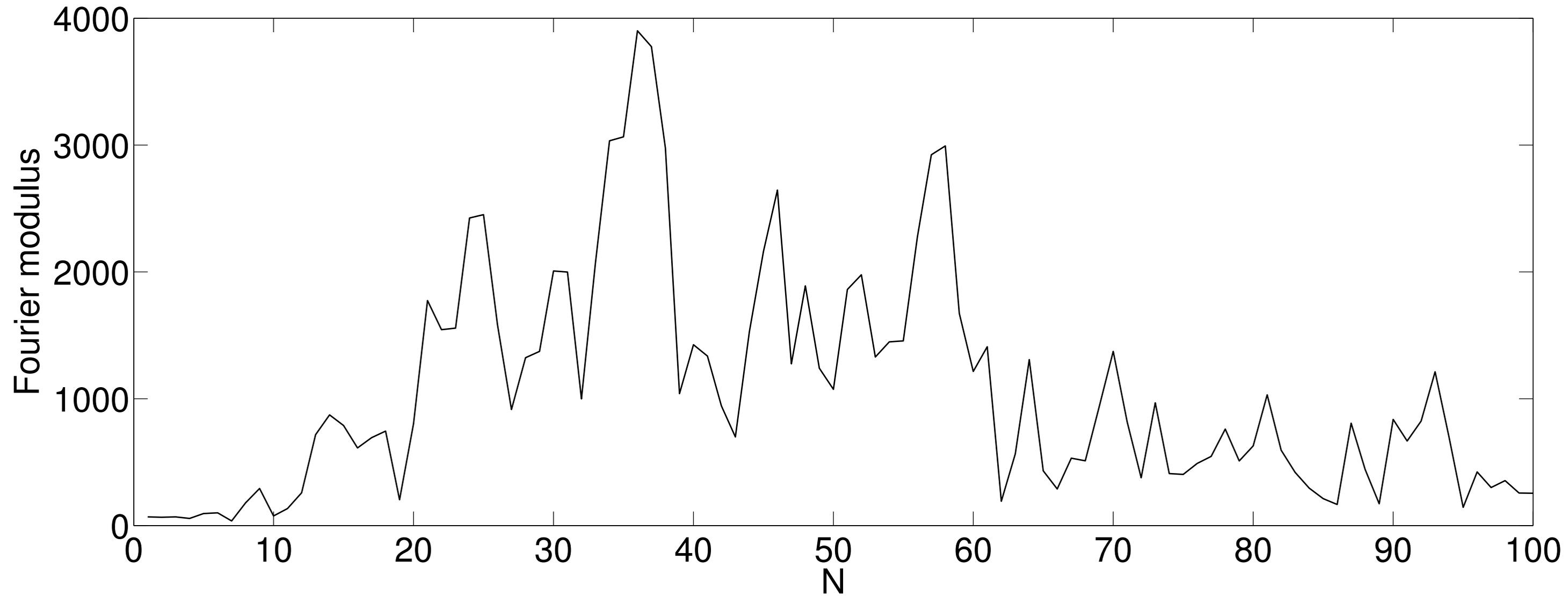

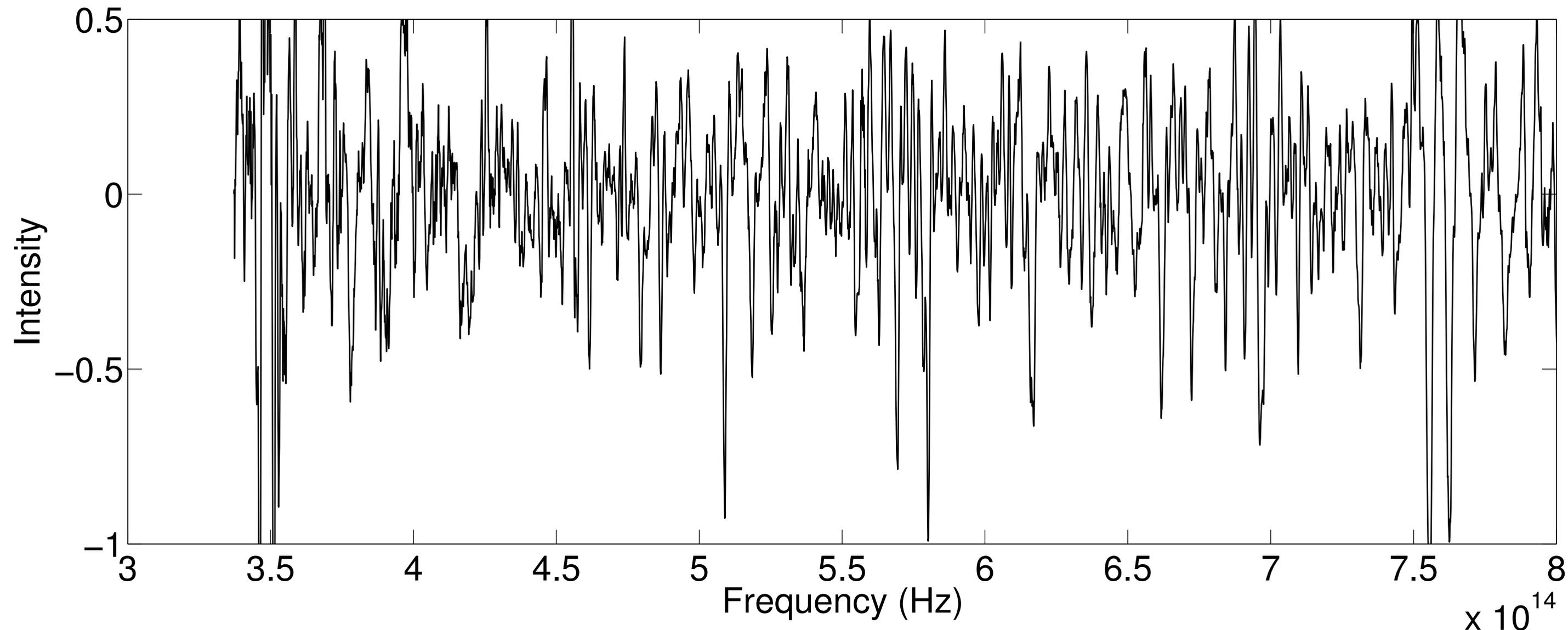

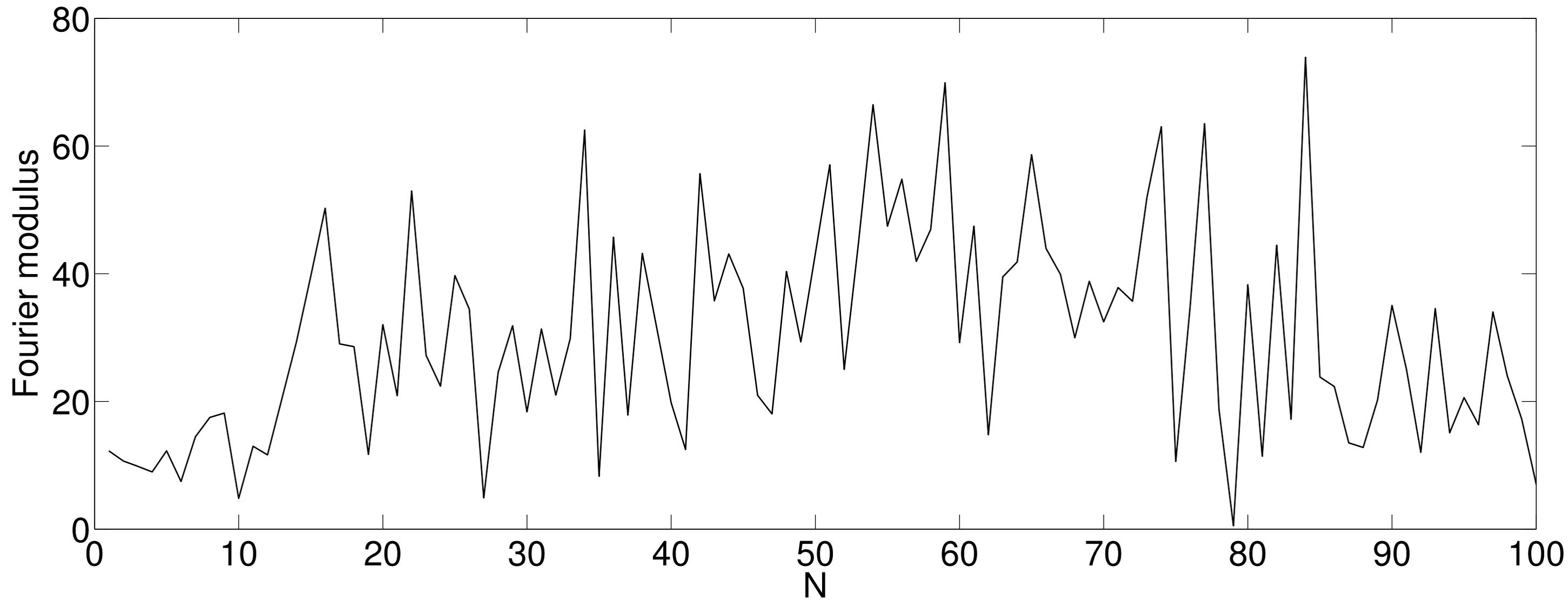

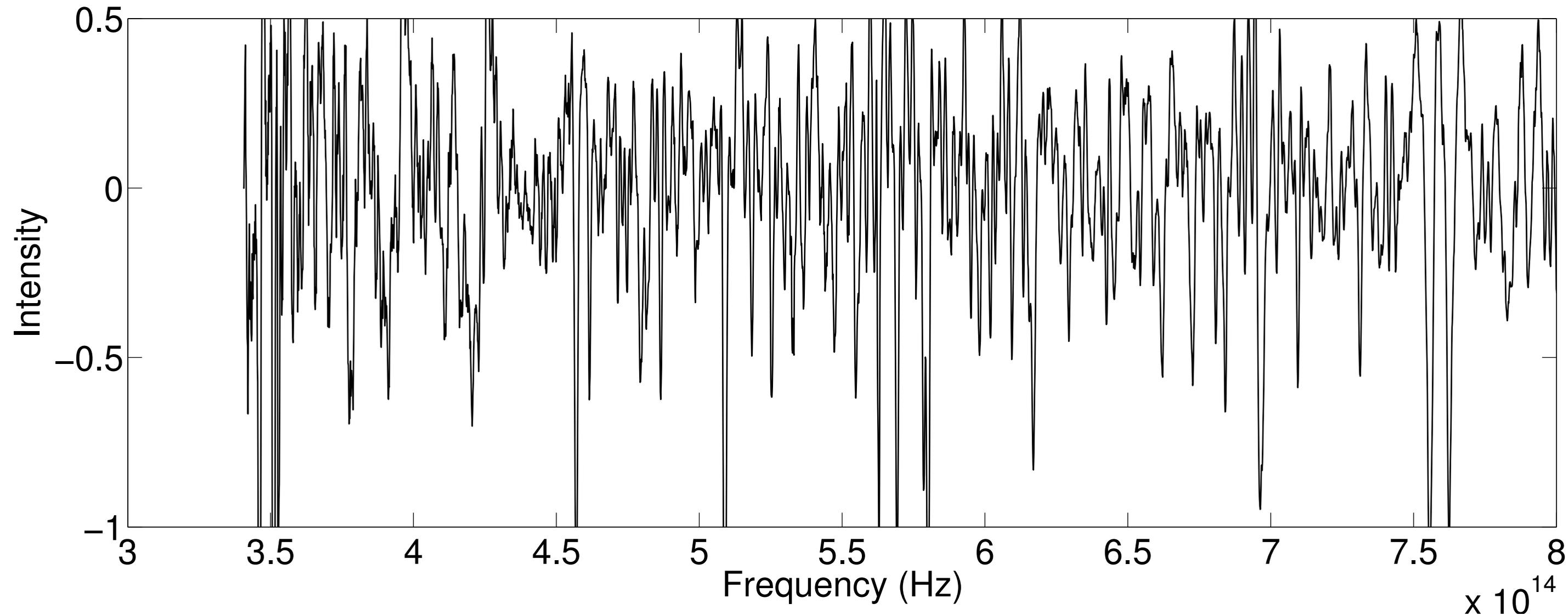

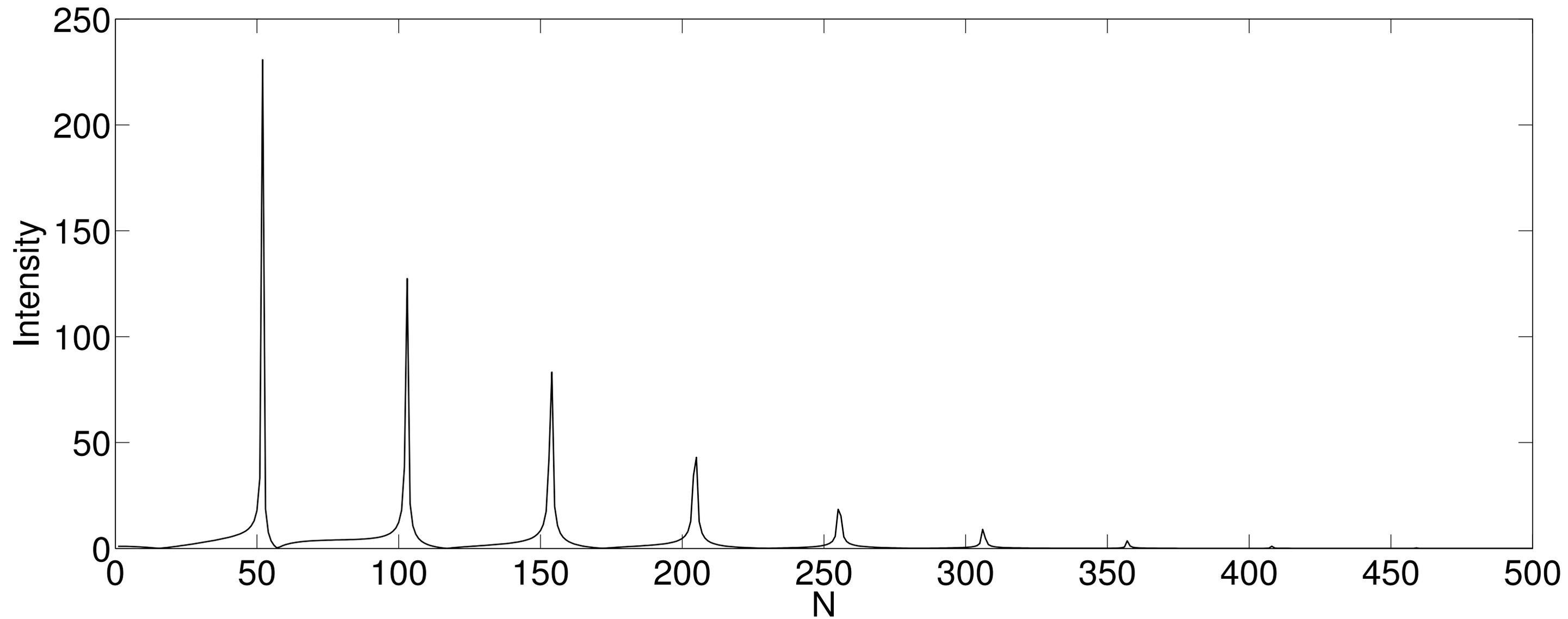

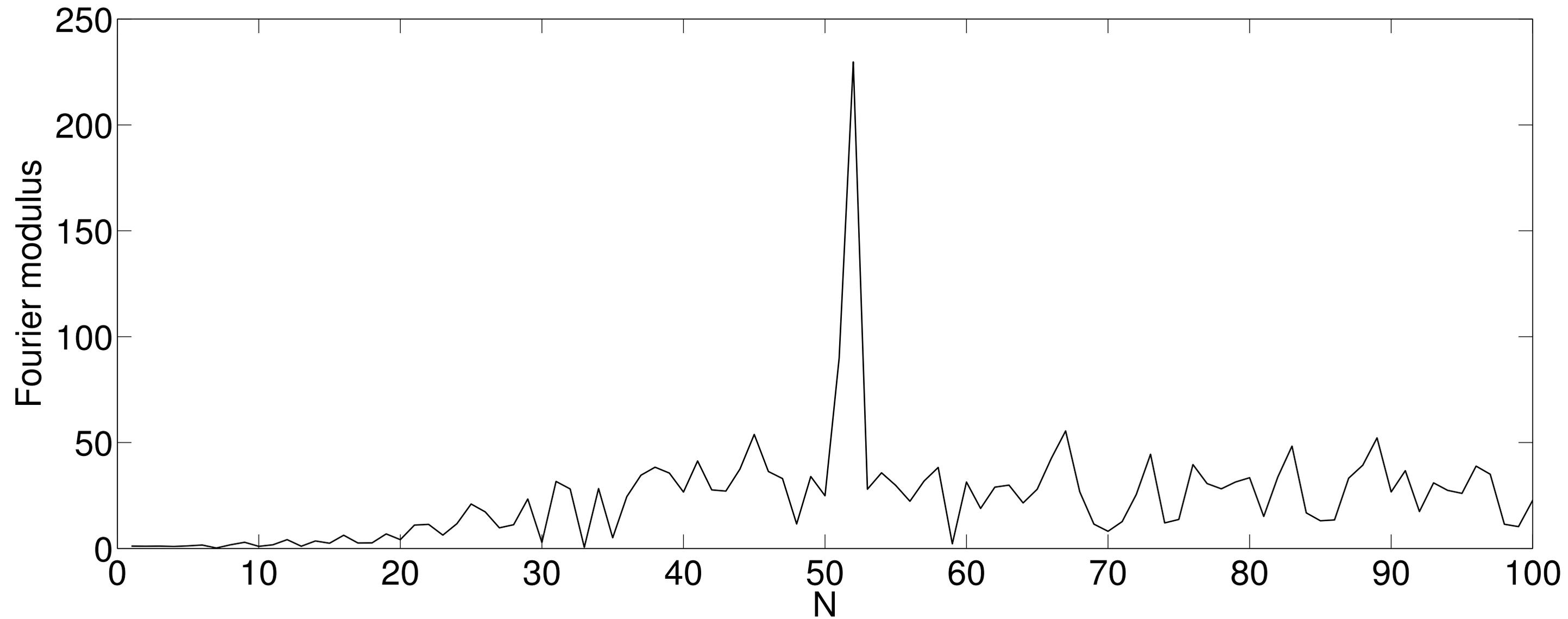

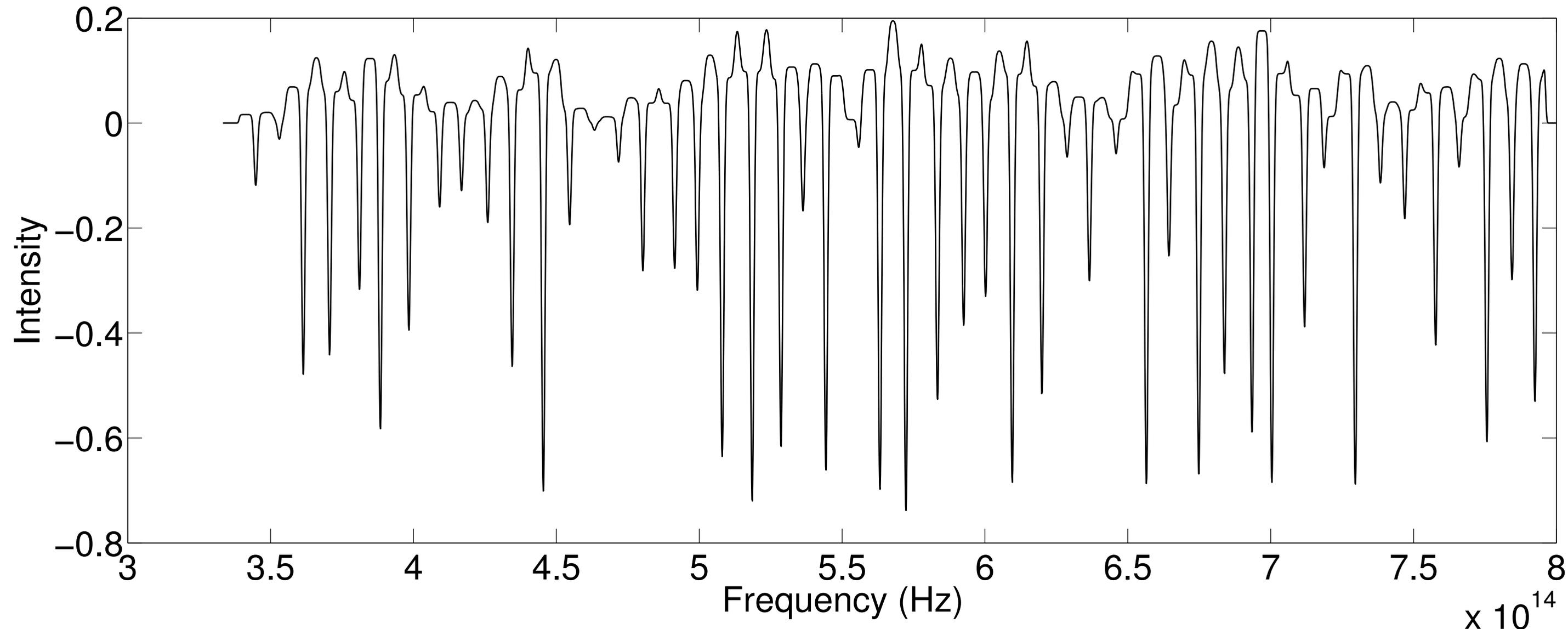